\documentclass[superscriptaddress,aps,pra,twocolumn,showpacs,nofootinbib,longbibliography,notitlepage]{revtex4-1}
\usepackage{etex}
\usepackage{amsmath,amssymb,amsthm}
\usepackage[colorlinks=true,citecolor=blue,urlcolor=blue]{hyperref}
\usepackage[pdftex]{graphicx}
\usepackage{times,txfonts}
\usepackage{braket}
\usepackage{color}
\usepackage{natbib}
\usepackage{amsmath,blkarray}
\usepackage{mathtools}
\usepackage{latexsym}
\usepackage{tabularx, booktabs}
\usepackage{graphics,epstopdf}
\usepackage{graphicx}
\usepackage{float}
\usepackage{graphicx}
\usepackage{amsfonts}
\usepackage{subcaption}
\usepackage{color,soul}

\newcommand{\be}{\begin{equation}}
\newcommand{\ee}{\end{equation}}
\newcommand{\ba}{\begin{eqnarray}}
\newcommand{\ea}{\end{eqnarray}}

\begin{document}
\title{Sharing preparation contextuality in Bell experiment by arbitrary pair of sequential observers  }
\author{ A. Kumari }
\email{asmita.physics@gmail.com}
\affiliation{S. N. Bose National Centre for Basic Sciences, Block JD, Sector III, Salt Lake, Kolkata 700106, India}
\author{ A. K. Pan }
\email{akp@phy.iith.ac.in}

\affiliation{Department of Physics, Indian Institute of Technology Hyderabad, Sangareddy, Telengana 502284, India }
\begin{abstract}
Based on the quantum violation of bipartite Bell inequality, it has been demonstrated that the sharing of non-locality can be demonstrated for at most two sequential observers at one end and at most one-pair of observers at both ends. In this work, we study the sharing of non-locality and preparation contextuality based on a bipartite Bell inequality, involving arbitrary  $n$ measurements by one party and $2^{n-1}$ measurements by other party. Such a Bell inequality has two bounds, the local bound and the preparation non-contextual bound, which is smaller than the local bound. We show that while non-locality can be shared only by first pair of the sequential observers, the preparation contextuality can be shared by arbitrary pair of independent sequential observers at both ends.
\end{abstract}
\maketitle
\section{Introduction}
\label{SecI}
Bell's theorem \cite{bell} is one of the most famous discoveries of quantum theory. It is the first no-go theorem that demonstrates that ontological model satisfying locality cannot reproduce all statistics of quantum theory. Such a feature is widely known as quantum non-locality which is demonstrated through the quantum violation of suitable Bell's inequalities. It is one of the elegant routes to demonstrate the supremacy of quantum theory over its classical counterpart. In addition to the extensive importance of Bell inequality in quantum foundations, it plays a pivotal role in quantum information processing like secure quantum key distribution \cite{bar05,acin06,acin07,pir09}, randomness certification \cite{col06,pir10,nieto,col12}, witnessing Hilbert space dimension \cite{wehner,gallego,ahrens,brunnerprl13,bowler,sik16prl,cong17,pan2020} and for achieving advantages in communication complexity tasks \cite{complx1}. 

There is another important no-go theorem-Kochen and Specker (KS) theorem \cite{kochen67} which provides a way of discriminating quantum theory from classical non-contextual theories. Over the years, many simpler proofs of KS theorem have been proposed  \cite{peres90,mermin93,cab08,pan10}. 
Importantly, contextuality can be revealed for a single system in contrast to non-locality which requires spatially separated entangled systems. However, the KS theorem only involves the measurement non-contextuality, and it is not applicable to unsharp measurement. Later, Spekkens \cite{spek05} generalized the notion of non-contextuality for arbitrary operational theories and extended the formulation to preparation, transformation and unsharp measurement. In recent times, the quantum preparation contextuality has been extensively studied  \cite{spek09,hameedi,ghorai18,saha19b,pan19,kumari2019,mukherjee22,schmid18,pan21}.

The aim of this work is to demonstrate the sharing of preparation contextuality by multiple independent sequential observers. Sharing of various forms of quantum correlations has gained considerable attention in recent times.   Based on the quantum violation of Clauser-Horne-Shimony-Halt (CHSH) inequality \cite{clauser69}  Silva \emph{et al.}\cite{silva2015} first demonstrated that at most two independent observers can share the non-locality sequentially when sharing is considered for one party. Let a bipartite two-qubit entangled state is shared by two distant observers (Alice and Bob) such that one qubit is in possession of Alice and the other qubit is with Bob who performs unsharp measurements and relays the qubit to the second Bob (next sequential observer), who does the same. Then, the sharing of non-locality implies that by recycling the qubit of sequential Bobs, the Bell inequality is violated by independent sequential observers. If two sequential Bobs violate Bell’s inequality, both of them cannot get the optimal quantum nonlocality rather they share the nonlocality. In a recent work, Brown and Colbeck  \cite{brown2020} demonstrated that the CHSH non-locality can be shared by an arbitrarily long sequence of independent observers. However, for every sequential observer, the unsharpness parameter and a new set of observables have to be chosen. This is in contrast to the earlier works \cite{silva2015}, where every sequential observer performs the same set of observables. Very recently, Cheng  \emph{et al.} \cite{Cheng2021,Cheng2022} showed that at most one pair of observers can share non-locality while considering the sharing for both parties.   In recent years, a flurry of works have been reported to examine the number of independent sequential observers sharing different quantum correlations, viz., entanglement \cite{bera2018}, steering \cite{sasmal2018,Akshata,Shashank}, non-locality \cite{silva2015,Karthik,sumit,brown2020,zh21,roy2020,Mao2022} and preparation contextuality \cite{kumari2019,anwer21}.

In this paper, we study the sharing of non-locality and preparation contextuality by multiple independent sequential observers at both ends. For this, we use a suitable Bell inequality proposed in \cite{ghorai18}  involving $n$ measurements by one party and $2^{n-1}$ measurements by other party. Such a Bell inequality has a local bound and a preparation non-contextual bound. The preparation non-contextual bound is considerably lower than the local bound. Such an inequality arises from a communication game known as parity-oblivious random access code (PORAC). We show that the sharing of non-locality is restricted only to single pair of observers, but sharing of preparation contextuality can be demonstrated for unbounded pair of sequential observers at both ends. For simplicity, we consider the symmetric case sharing when $k^{th}$ observer at both ends can share preparation contextuality.

This paper is organized as follows. In Sec. II we demonstrate the PORAC game which is used as a tool to demonstrate our results. In Sec. III, we discuss the Bell expression and its local, preparation non-contextual and quantum bounds. In Sec. IV, we derive the sequential quantum value of the Bell expression. The condition of sharing of non-locality and preparation contextuality at both ends are demonstrated in Sec. V and Sec. VI respectively. In Sec. VII we provide the analytical proof of sharing by unbounded pair of sequential observers at both ends. Finally, in Sec VIII, we discuss our results.

\section{Arbitrary input Bell inequality and it's local, preparation non-contextual and quantum bound}
We briefly discuss the notion of PORAC which is a two-party one-way communication game, used as a tool to demonstrate our results. A $n$-bit RAC \cite{spek09,ambainis} involves a sender Alice and a receiver Bob. Alice has $n$-bit string $x$  chosen randomly from $x\in\{0,1\}^{n}$ and Bob receive inputs $ y \in \{1,2, ..., n\}$ uniformly at random and outputs $b$. Bob's task is to  recover ${y}^{th}$ bit of Alice input with a probability, i.e., the winning condition of the game is $b=x_y$. The average success probability of the game is defined as,
	\begin{equation}
		\label{qprob}
		P = \dfrac{1}{2^n n}\sum\limits_{x,y}p(b=x_y|x,y).
	\end{equation}
The task of Alice and Bob is to maximize the success probability. To help Bob, Alice can communicate some information with him. However, there is a parity-oblivious condition which dictates that Alice can communicate any number of bits to Bob but no information about the parity of $x$ should be transmitted.

Spekkens \emph{et al.} \cite{spek09} defined the parity-oblivious condition with respect to a parity set $ \mathbb{P}_n= \{x|x \in \{0,1\}^n,\sum_{r} x_{r} \geq 2\} $ with $r\in \{1,2,...,n\}$. For any $s \in \mathbb{P}_{n}$, no information about $s\cdot x = \oplus_{r} s_{r}x_{r}$ (s-parity) is to be transmitted to Bob, where $\oplus$ is sum modulo $ 2 $. The maximum average success probability in such a classical $ n $-bit PORAC is \cite{spek09}

	\begin{align}
		\label{cb}
	P_{pnc}	 \leq \frac{1}{n} + \frac{(n-1)}{2n} = \frac{1}{2}\left(1+\frac{1}{n}\right) 
	\end{align}
 
It has been demonstrated in \cite{,spek05,spek09,pan19,pan21} that the parity-obliviousness at the operational level must be satisfied at the level of ontic states if the ontological model of quantum theory is preparation non-contextual. Thus, in a preparation non-contextual model, the classical bound remains the same as given in Eq.(\ref{cb}). Quantum violation of this bound thus demonstrates a form of non-classicality - the preparation contextuality. Throughout this paper, by quantum preparation contextuality, we refer to the violation of the preparation non-contextuality inequality in Eq.(\ref{cb}).

In quantum mechanics, Alice encodes her $ n $-bit string of $ x \in \{1,2,..., 2^{n}\}$ into quantum states $ \rho_x $. Here we consider entanglement assisted version of PORAC  so that Alice and Bob share a suitable entangled state. By performing $2^{n-1}$ dichotomic measurement, Alice can steer the state $\rho_x$ to Bob. After receiving the state  $ \rho_x $, for every $y \in \{1,2,...,n\}$, Bob performs a dichotomic measurement and reports the outcome $b$ as his output.	
The quantum success probability can be written as \cite{ghorai18}, 
\begin{align}
\label{qprobn}
p_Q = \dfrac{1}{2} + \dfrac{1}{2^n n}\sum_{y=1}^{n} \sum\limits_{i=1}^{2^{n-1}} (-1)^{x^i_y}  \langle A_i\otimes B_y\rangle 
\end{align}
It is seen from Eq.(\ref{qprobn}) that, the quantum success probability is dependent solely on the Bell expression 
\begin{align}
\label{nbell1}
	\mathcal{B}_{n} =  \sum_{y=1}^{n}\sum_{i=1}^{2^{n-1}} (-1)^{x^i_y}  A_{n,i}\otimes B_{n,y} 
\end{align}
where $n$ is arbitrary integer with $n>2$. For $n=2$ and $3$, the Bell expressions $\mathcal{B}_{n}$ become the well-known CHSH \cite{clauser69} and Gisin's elegant Bell \cite{gisin} expressions. Using an elegant sum-of-squares decomposition \cite{ghorai18} the optimal quantum value of the Bell expression was derived as \cite{pan2020} 
\begin{align}
\label{maxq}
{(\mathcal{B}^{opt}_{n})}_{Q}  = 2^{n-1}\sqrt{n}
\end{align}
This is achieved when Alice and Bob share $\lfloor\frac{n-1}{2}\rfloor$ number of two-qubit maximally entangled states and Bob performs the measurements of $n$ number of mutually anti-commuting observables. Then, the optimal quantum success probability for an $n$-bit PORAC is
\begin{align}
P^{opt}_Q =\frac{1}{2}\bigg( 1+\frac{1}{\sqrt{n}} \bigg)
\end{align}
Since for any $n$, $p^{opt}_Q > P^{opt}_{pnc} = \frac{1}{2}\left(1+\frac{1}{n}\right)$, the quantum preparation contextuality is demonstrated. In other words, using quantum resources, the success probability of the $n$-bit PORAC exceeds the preparation non-contextual bound. In this paper, we will use the Bell expression in Eq.(\ref {nbell1}) to demonstrate the sharing of non-locality and preparation contextuality at both ends.

The local bound of the Bell expression in Eq.(\ref {nbell1}) is derived  as \cite{kumari2019}
\begin{align} 
\label{localbound}
	{(\mathcal{B}_{n})}_{L} \leq n\binom{n-1}{\lfloor\frac{n-1}{2}\rfloor}  
\end{align}
For $n=2$ and $3$, we have,  ${(\mathcal{B}_{n})}_{L}\leq 2$ (CHSH inequality) and ${(\mathcal{B}_{n})}_{L} \leq 6$ (Gisin's elegant Bell inequality \cite{gisin}) respectively.

Importantly, the optimal quantum value derived in Eq.(\ref{maxq}) automatically satisfy the parity oblivious condition in Eq.(\ref{poc}) \cite{ghorai18}. In quantum theory, the parity-oblivious condition demands that 
	\begin{align}
		\label{poc}
		\forall s: \frac{1}{2^{n-1}}\sum\limits_{x|x.s=0} \rho_{x}=\frac{1}{2^{n-1}}\sum\limits_{x|x.s=1} \rho_{x}
	\end{align}
	
For $n$-bit case, total  $C_n= 2^{n-1} - n$ non-trivial parity-obliviousness condition between Alice's observables \cite{ghorai18} needs to satisfy 
\begin{equation}
\label{ntpnc}
\forall s: \sum_{i=1}^{2^{n-1}} (-1)^{s.x^i} A_{n,i} = 0
\end{equation}
 The parity-oblivious conditions in Eq.(\ref{ntpnc}) will have equivalent representation in an ontological model, which is in the premise of the preparation non-contextuality assumption in an ontological model. Putting the condition in Eq.(\ref{ntpnc}) to Eq.(\ref{nbell1}), we have \cite{ghorai18}
\begin{align}
\label{npnc}
	(\mathcal{B}_{n})_{pnc} \leq  2^{n-1}
\end{align}
which is preparation non-contextual bound of the Bell expression in Eq.(\ref {nbell1}).
Comparing local bound in Eq.(\ref{localbound}) and  preparation non-contextual bound in Eq.(\ref{npnc}) we can see that $(\mathcal{B}_{n})_{L}> 	(\mathcal{B}_{n})_{pnc}$. Then there may be instances when non-locality may not be revealed but a non-classicality in the form of preparation contextuality can be revealed. This indicate that there is a higher chance of sharing the preparation contextuality for more pairs of sequential observers than non-locality. In this work, we examine the maximum number of pairs of Alice and Bob can share the non-locality and the preparation contextuality across both Alice's and Bob's ends based on the quantum violations of Eq.(\ref{localbound}) and Eq.(\ref{npnc}) respectively.

 \section{Sequential Quantum value of the Bell expression}
We start by pointing out that the sharing of any quantum correlation protocol requires the prior observers to perform unsharp measurements \cite{busch}  represented by a set of positive operator valued measures (POVMs). Although sharp measurement seems advantageous from information-theoretic perspective, there remain certain tasks where unsharp measurement showcases its supremacy over sharp measurement \cite{std1,std2,ran1}. An ideal sharp measurement extracts maximum information by collapsing the system state into  one of the eigenstates of the measured observable. This causes maximum disturbance to the initial system. For an entangled state if sharp measurement is performed by one party the system become a mixed state. No quantum violation of Bell's inequality can be found by sequential observer. However, if previous observers perform unsharp measurement, the initial entangled state is partially disturbed. This can be controlled by tuning the degree of unshrpness parameter. In such a case some amount of entanglement may be remained in the system, which can be used by sequential observers. Then, by using this residual entanglement there remains a chance of sharing non-locality through the quantum violation of Bell's inequality.
 
 In order to demonstrate sharing of non-locality and preparation contextuality, we consider that an entangled state is shared between multiple independent observers. For the Bell expression $\mathcal{B}_{n}$ in Eq.(\ref {nbell1}), each sequential Alice and Bob perform dichotomic measurements upon receiving input $x \in \{1,2,...,2^{n-1} \} $ and $y \in \{,2,...,n\}$ respectively. Here it should be noted that Alice's and Bob’s choices of measurement settings are completely random and in each run, they use the same set of their respective measurement settings. For any arbitrary  $n$, the respective POVMs that are performed by $k$ number of sequential Alices and  $l$ number of sequential Bobs are represented by 
\begin{align}
{E}^{\pm}_{n,x,p}=    \frac{1  \pm \eta_{n,p}}{2} \Pi^{+}_{A_{n,x,p}}  + \frac{1  \mp \eta_{n,p}}{2} \Pi^{-}_{A_{n,x,p}}
\end{align}
and
\begin{align}
{E}^{\pm}_{n,y,q}=    \frac{1  \pm \chi_{n,q}}{2} \Pi^{+}_{B_{n,y,q}}  + \frac{1  \mp \chi_{n,q}}{2} \Pi^{-}_{B_{n,y,q}}
\end{align}
with $p = (1,2,...k)$ and  $q = (1,2,...l)$. Here $\eta_{n,p}$ ($ \chi_{n,q}$) is the sharpness parameters of $p$th Alice ($q$th Bob) satisfying $0 \leq \eta_{n,p}, \chi_{n,q} \ \leq 1$. Then, the post-measurement state after unsharp measurement of $(k-1)^{th}$ Alice (Alice$_{k-1}$) and unsharp measurement of $(l-1)^{th}$ Bob (Bob$_{l-1}$) can be written as
\begin{widetext}
\begin{eqnarray}
\label{fbell}
\rho_{n,k,l} &=& \frac{1}{ 2^{n-1} n}\sum_{a,b \in \left\{ + ,- \right\}} \sum_{x = 1}^{2^{n-1}}\sum_{y = 1}^{n} \bigg[\bigg(\sqrt{{E}^{a}_{n,x,k-1}} \otimes \sqrt{{E}^{b}_{n,y,l-1}}\bigg) \rho_{n,k-1,l-1} \bigg(\sqrt{ {E}^{a}_{n,x,k-1}} \otimes \sqrt{ {E}^{b}_{n,y,l-1}} \bigg)\bigg]\\ \nonumber &=& \sqrt{1- \eta_{n,k-1}^2} \sigma_{n,k-1,l-1} + \frac{1-\sqrt{1- \eta_{n,k-1}^2}}{2^{n-1}}\sum_{a \in \left\{ + ,- \right\}}\sum_{x = 1}^{2^{n-1}} ( \Pi^{a}_{A_{n,x,k-1}} \otimes \mathbb{I} ) \sigma_{n,k-1,l-1} (\Pi^{a}_{A_{n,x,k-1}} \otimes \mathbb{I})
\end{eqnarray}

where, $ \sigma_{n,k-1,l-1}  $ is the reduced state after unsharp measurement of $(l-1)^{th}$ Bob is given by
\begin{eqnarray}
 \sigma_{n,k-1,l-1} &=& \sqrt{1-\chi_{n,l-1}^2}  \rho_{n,k-1,l-1} + \frac{1-\sqrt{1-\chi_{n,l-1}^2}}{n}\sum_{b \in \left\{ + ,- \right\}}\sum_{y = 1}^{n} ( \mathbb{I} \otimes {\Pi}^{b}_{B_{n,y,l-1}}   ) \rho_{n,k-1,l-1} ( \mathbb{I} \otimes {\Pi}^{b}_{B_{n,y,l-1}}  )
\end{eqnarray}
\end{widetext}
If the initial system state shared by Alice$_1$ and  Bob$_1$ is maximally entangled state as mentioned earlier,  the quantum value of the Bell expression given in Eq.(\ref{nbell1}) for Alice$_{k}$ and Bob$_l$ is obtained as
\begin{eqnarray}
\label{quant}
(\mathcal{B}_n^{k,l})_Q &=& 2^{n-1}\sqrt{n} \prod^{k-1}_{p=1} \gamma^A_p \prod^{l-1}_{q=1} \gamma^B_q \eta_{n,k} \chi_{n,l}
\end{eqnarray}
where, $\gamma^{A}_{n,p} = \frac{ \bigg(1+(n-1)\sqrt{1-\eta_{n,p}^2} \bigg)}{n} $ and $\gamma^{B}_{n,q} = \frac{ \bigg(1+(n-1)\sqrt{1-\chi_{n,q}^2} \bigg)}{n} $.
Now, the non-locality and the preparation contextuality can be shared by Alice$_k$ and Bob$_l$, if $(\mathcal{B}^{k,l}_n)_Q>(\mathcal{B}_n)_{L}$ and  $(\mathcal{B}^{k,l}_n)_Q>(\mathcal{B}_n)_{pnc}$ are respectively satisfied. \\

\section{Sharing non-locality with equal number of sequential Alice and  Bob}
We first examine how many sequentially independent pairs of Alices and Bobs can share non-locality through the quantum violation of ${(\mathcal{B}_{n})}_{L} $ in Eq.(\ref{localbound}). In order to share non-locality by Alice$_k$ and Bob$_l$, $(\mathcal{B}^{k,l})_Q  > n\binom{n-1}{\lfloor\frac{n-1}{2}\rfloor}$ should be satisfied. For the case of $n=2$, Bell inequality in  Eq.(\ref{localbound}) reduces to CHSH inequality. 
The quantum expression in Eq.(\ref{fbell}) shared between  Alice$_1$ and Bob$_1$ reduce to $ (\mathcal{B}_2^{1,1})_Q = 2 \sqrt{2} \eta_{2,1} \chi_{2,1}$. Then quantum violation of Bell inequality by Alice$_1$ and Bob$_1$ requires $\eta_{2,1} =  \chi_{2,1} \geq 0.841$ and the maximum violation of $2 \sqrt{2}$ is obtained for $\eta_{2,1} =  \chi_{2,1} = 1$. The quantum value if CHSH  expression for Alice$_2$ and Bob$_2$ is obtained as
 \begin{eqnarray}
 \label{a2b2}
 \nonumber
 (\mathcal{B}_2^{2,2})_Q = \frac{2 \sqrt{2}}{4} \bigg[ \bigg(1+ \sqrt{1-\eta^2_{2,1}}\bigg) \bigg(1+ \sqrt{1-\chi^2_{2,1}}\bigg) \eta_{2,2} \ \ \chi_{2,2}\bigg] \\
\end{eqnarray}
Which is dependent on the sharpness parameters of Alice$_1$ and Bob$_1$. By considering the critical values $\eta^*_{2,1} =  \chi^*_{2,1} = 0.841$, such that  Alice$_1$ and Bob$_1$ just violate the Bell inequality, the maximum quantum value for Alice$_2$ and Bob$_2$ is obtained to be $1.679$ when $\eta_{2,2} =  \chi_{2,2} = 1$. It approaches $2$ for either or both of $\eta_{2,2}$ and $\chi_{2,2}$ greater than $1$. Hence, within the valid range of sharpness parameters  the non-locality cannot be shared by Alice$_2$ and Bob$_2$. This result is already studied in \cite{Cheng2021,Cheng2022}.\\

\begin{figure}[htp]
    \includegraphics[width=8cm]{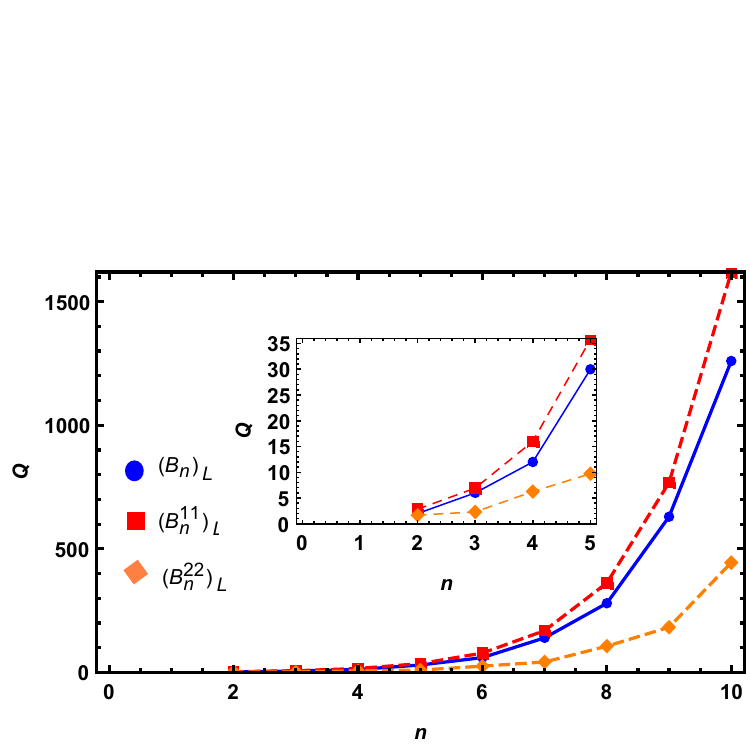}
    \caption{ The local bound (blue circle on continuous line)  and maximum quantum value obtained by  Alice$_1$-Bob$_1$  (  red square on dashed line) and  Alice$_2$-Bob$_2$  (  orange diamond shape on dashed line) up to $n=10$ bits.}
    \label{fig:BLP}
\end{figure}

For $n=3$, Bell inequality in Eq.(\ref{localbound}) becomes Gisin's elegant Bell inequality with local bound $6$. The quantum expression of Bell inequality for Alice$_1$ and Bob$_1$ is obtained as $ 4 \sqrt{3} \eta_{3,1} \chi_{3,1} $. The quantum violation of Bell inequality by Alice$_1$ and Bob$_1$ requires $\eta_{3,1} =  \chi_{3,1} \geq 0.931$ and the maximum violation of $4 \sqrt{3}$ is obtained for $\eta_{2,1} =  \chi_{2,1} = 1$. The quantum value of elegant Bell expression for Alice$_2$ and Bob$_2$ is
 \begin{eqnarray}
 \label{a3b3}
 \nonumber
 (\mathcal{B}_3^{2,2})_Q = \frac{4 \sqrt{3}}{9} \bigg[ \bigg(1+ 2 \sqrt{1-\eta^2_{3,1}}\bigg) \bigg(1+ \sqrt{1- 2 \chi^2_{3,1}}\bigg) \eta_{3,2} \ \ \chi_{3,2}\bigg] \\
\end{eqnarray}
By considering the critical values $\eta^*_{3,1} =  \chi^*_{3,1} = 0.931$, the maximum quantum value for Alice$_2$ and Bob$_2$ is obtained to be $2.304$ at $\eta_{3,2} =  \chi_{3,2} = 1$, which is much less than the local bound $6$.  Hence also for the case of $n=3$, the non-locality cannot be shared by Alice$_2$ and Bob$_2$.\\

Although it is not expected that more than one pair of Alice and Bob can share the non-locality for large $n$ but for completeness we examine higher value of $n$. In Fig. 1 we have plotted the local bound ($(\mathcal{B}_n)_L $, blue circle on continuous line) and maximum quantum values obtained by  Alice$_1$-Bob$_1$ ( $(\mathcal{B}_n^{11})_Q $, red square on dashed line) and  Alice$_2$-Bob$_2$ ( $(\mathcal{B}_n^{22})_Q $, orange diamond shape on dashed line) up to $n=10$. From Fig.1, it can be seen that the maximum quantum value of $(\mathcal{B}_n^{2,2})_Q $ for  Alice$_2$-Bob$_2$ ( orange diamond shape ) always remains less than the respective local bound $(\mathcal{B}_n)_{L} $ (blue circle ). It can also be seen that the difference between $(\mathcal{B}_n)_{L} $ blue circle )  and $(\mathcal{B}_n^{2,2})_Q $  ( orange diamond shape ) increases as $n$ increases. We then can say that if both Alice and Bob perform unsharp measurement, using the violation of local bound of Bell expression $\mathcal{B}_{n}$ in Eq.(\ref {nbell1}), the possibility of sharing non-locality by second pair of Alice and Bob decreases as $n$ increases, i.e., sharing of non-locality remain restricted to one pair of sequential observers as claimed in \cite{Cheng2021,Cheng2022}.  
Now, since $ (\mathcal{B}_n)_{L} >(\mathcal{B}_{n})_{pnc} $, we may expect the sharing of preparation contextuality can be demonstrated by more pair of sequential Alices and Bobs. In next section, we have studied this through the violation of preparation non-contextual inequality in Eq.(\ref{npnc}).

\section{Sharing preparation contextuality }
Note that for the case of $n=2$, the  Bell inequality  in Eq.(\ref{nbell1})  reduces to CHSH inequality for which both the local and preparation non-contextual bound is the same. Hence, the analysis for $n=2$, remains the same as we did for sharing the non-locality. As mentioned, the Bell inequality in Eq.(\ref{nbell1}) has two classical bounds, the local bound and preparation non-contextual bound, but quantum bound remains same. The quantum expression for Alice$_1$ and Bob$_1$ for $n=3$ is again  $ 4 \sqrt{3} \eta_{3,1} \chi_{3,1} $. The quantum violation of the preparation non-contextuality bound $(\mathcal{B}_{n})_{pnc} \leq 4$ requires, $\eta_{3,1} =  \chi_{3,1} \geq 0.840$ which is less than what is  required for sharing the non-locality. For the case of  Alice$_2$ and Bob$_2$, the quantum expression of Bell inequality is given in Eq.(\ref{a2b2}). Then by considering critical values  $\eta^*_{3,1} =  \chi^*_{3,1} = 0.840$, such that Alice$_1$ and Bob$_1$ just violates the Bell inequality, the maximum quantum value of the Bell expression for Alice$_2$ and Bob$_2$ is obtained to be $4.073$ at $\eta_{3,2} =  \chi_{3,2} = 1$. Thus, for $n=3$, sharing of the preparation contextuality can be demonstrated for the second pair of sequential observers.

Next, the quantum value of elegant Bell expression for Alice$_3$ and Bob$_3$ can be written as
 \begin{eqnarray}
 \label{a4b4}
 \nonumber
 (\mathcal{B}_3^{3,3})_Q &=& \frac{4 \sqrt{3}}{81} \bigg[ \bigg(1+ 2 \sqrt{1-\eta^2_{3,1}}\bigg)\bigg(1+ 2 \sqrt{1-\eta^2_{3,2}}\bigg)\\ & \times & \bigg(1+ \sqrt{1- 2 \chi^2_{3,1}}\bigg)\bigg(1+ \sqrt{1- 2 \chi^2_{3,2}}\bigg) \eta_{3,3} \ \chi_{3,3}\bigg] 
\end{eqnarray}
which depends on the sharpness parameter of all the previous observers Alice$_1$, Bob$_1$, Alice$_2$ and Bob$_2$. Substituting the critical values $\eta^*_{3,1} =  \chi^*_{3,1} = 0.840$ and $\eta^*_{3,2} =  \chi^*_{3,2}  = 9.742$, the maximum quantum value of $(\mathcal{B}_3^{3,3})_Q$ is obtained to be $0.727 $ at $\eta_{3,3} =  \chi_{3,3}  = 1$. Hence for $n=3$ only two pairs of Alices and Bobs can share preparation contextuality.\\

Further, we have numerically studied for $n=4$ and found the quantum value for Alice$_3$ and Bob$_3$ increases with $n$, but cannot beat preparation non-contextuality bound. From Fig. 2, it can be seen that the sharing of preparation contextuality is restricted for Alice$_2$ and Bob$_2$ up to $n=7$. For $n=8$, the preparation non-contextuality bound is $128$ and the maximum quantum value obtained for Alice$_3$ and Bob$_3$ is $134.012$, thereby Alice$_3$ and Bob$_3$ can also share the preparation contextuality. In Fig. 2, we have shown the result up to $n=10$. However, for a higher value of $n$, the sharing of preparation contextuality can be demonstrated for more number of sequential Alices and Bobs. We analytically show that for a sufficiently large value of $n$, an unbounded number of Alices and Bobs can share preparation contextuality. \\

\begin{figure}[htp]
  \includegraphics[width=8cm]{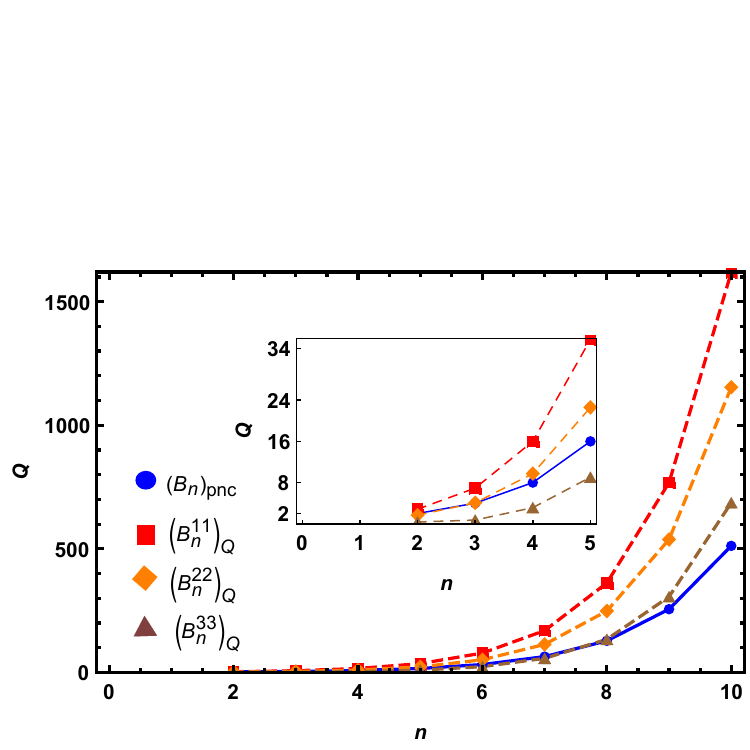}
\caption{ Preparation non-contextuality bound (blue circle on continuous line) ) and the maximum quantum value obtained by  Alice$_1$-Bob$_1$ (  red square on dashed line),  Alice$_2$-Bob$_2$ ( orange diamond shape on dashed line) and  Alice$_3$-Bob$_3$ (brown triangle on dashed line) upto $n=10$ bits.}
    \label{fig:BLP}
\end{figure}
\section{Generalization of sharing non-trivial preparation contextuality }
In order to show the sharing of preparation contextuality by arbitrary pair of sequential observers, from Eq.(\ref{quant}), we write the general condition on the sharpness parameters for Alice$_k$ and Bob$_l$ for sequentially sharing preparation contextuality is given by
\begin{eqnarray}
 \eta_{n,k} \chi_{n,l} > \frac{1}{\sqrt{n} \prod^{k-1}_{p=1} \gamma^A_{n,p} \prod^{l-1}_{q=1} \gamma^B_{n,q} } 
\end{eqnarray}
For convenience let us assume that $l=k$, and the unsharpness parameters of Alice$_k$ and Bob$_k$ for a given $k$ are equal, i.e., $\chi_{n,l} = \eta_{n,k} $. Then the condition on the unsharpness parameter for sharing preparation contextuality becomes
\begin{eqnarray}
\label{eta}
 \eta^2_{n,k}  > \frac{1}{\sqrt{n} \bigg(\prod^{k-1}_{p=1} \gamma_{n,p} \bigg)^2  } 
\end{eqnarray}
Using Eq.(\ref{eta}), the critical value of the sharpness parameter of Alice$_k$ and Bob$_k $ above which violation of non-contextual inequality in Eq.(\ref{npnc}) is obtained is given by
\begin{eqnarray}
\label{eta1}
 \eta^2_{n,k}  = \frac{1}{\sqrt{n} \bigg(\gamma_{n,k-1} \prod^{k-2}_{p=1} \gamma_{n,p} \bigg)^2  } 
\end{eqnarray}
Then, in order to share the contextuality by Alice$_k$ and Bob$_k$, the sharpness parameter requires 
\begin{eqnarray}
\label{eta0}
 \eta^2_{n,k}  \geq \frac{1}{\sqrt{n} \bigg(\prod^{k-1}_{p=1} \gamma_{n,p} \bigg)^2  } 
\end{eqnarray}
Now, for Alice$_{k-1}$ and Bob$_{k-1}$, the critical value of sharpness parameter is
\begin{eqnarray}
\label{eta2}
 \eta^2_{n,k-1}  = \frac{1}{\sqrt{n} \bigg( \prod^{k-2}_{p=1} \gamma_{n,p} \bigg)^2  } 
\end{eqnarray}
\begin{figure}[htp]
    \centering
    \includegraphics[width=9cm]{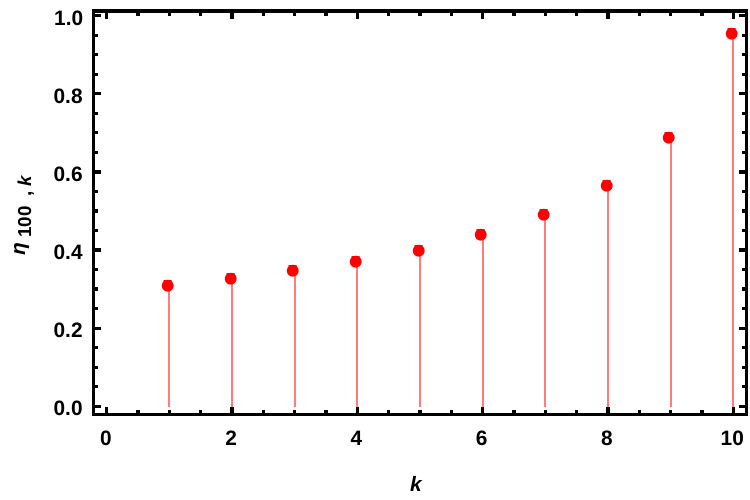}
    \caption{ The minimum value of sharpness parameter of $10$ pairs of Alice and Bob required for violating the preparation non-contextual bound for the family of Bell's expression for $n=100$. }
    \label{fig:3}
\end{figure}
Using Eq.(\ref{eta2}) the condition on sharpness parameter for sharing non-trivial preparation contextuality for Alice$_{k}$ and Bob$_{k}$ in Eq.(\ref{eta0}) reduces to
\begin{eqnarray}
\label{eta3}
 \eta_{n,k}  \geq \frac{\eta_{n,k-1}}{ \gamma_{n,k-1}    } 
\end{eqnarray}
Since,  $\gamma_{n,k-1}  = \frac{ \bigg(1+(n-1)\sqrt{1-\eta_{n,p}^2} \bigg)}{n} > \sqrt{1-\eta_{n,k-1}^2} $ , we can rewrite  Eq.(\ref{eta3}) as
\begin{eqnarray}
\label{eta4}
 \eta_{n,k}  \geq \frac{\eta_{n,k-1}}{ \sqrt{ 1-\eta_{n,k-1}^2}  } 
\end{eqnarray}
Here, it should be noted that the critical value of unsharpness parameter for Alice$_1$ and  Bob$_1$ requires, $\eta^{*2}_{n,1} = 1/ \sqrt{n} $, i.e.,  $\eta^*_{n,1} = (1/n)^{1/4} $. Using Eq.(\ref{eta4}), we have the critical value of unsharpness parameter for Alice$_2$ and Bob$_2$ as $\eta_{n,2}   = \frac{1}{\sqrt{\sqrt{n}-1}}$ and for Alice$_3$ and Bob$_3$, $\eta_{n,3} =  \frac{1}{\sqrt{\sqrt{n}-2}}$.
Thus, the unsharpness parameter for Alice$_k$ and Bob$_k$ have to satisfy
\begin{eqnarray}
\label{eta7}
 \eta_{n,k} \geq  \frac{1}{\sqrt{\sqrt{n}-(k-1)}}
\end{eqnarray}
In other words, preparation contextuality can be shared by arbitrary $k$ pairs of Alice and Bob if Eq.(\ref{eta7}) is satisfied. If final pair of Alice and Bob perform the sharp measurement, we have the condition
\begin{eqnarray}
\label{eta9}
n(k)  \geq  k^2
\end{eqnarray}
as $k$ is very large. Hence, using the Bell expression for  $n=k^2$ the preparation contextuality can be shared by $k$ independent sequential pair of Alice and Bob.
In Fig.3 we have shown that using quantum violation of the Bell's inequality for $n=100$ at most $10$ pairs of Alices and Bobs can sequentially share  preparation contextuality. 

\section{Summary and Discussion}
In summary, we examined the maximum number of pairs of Alice and Bob that can share the non-locality and the preparation contextuality. Our study is based on the quantum violation of suitable bipartite Bell inequality, arising from a communication game known as parity-oblivious random-access-code. Such a Bell inequality involves arbitrary  $n$ measurements by one party and $2^{n-1}$ measurements by the other party. As mentioned, it has two different classical bounds, the local bound and the preparation non-contextual bound which is lower than the local bound for $n>2$. 

We demonstrated that if the sharing is considered for both parties then at most one pair of Alice and Bob can share the non-locality irrespective of the value of $n$. This result is in accordance with the previous works \cite{Cheng2021,Cheng2022} who considered $n=2$ case.  Since the preparation non-contextual bound is lower than the local bound and the optimal quantum value remains the same, there may be more pairs that can share the preparation contextuality.  Indeed, we have shown that the preparation contextuality can be shared by an arbitrary pair of independent sequential observers at both ends of the bipartite Bell experiment for a sufficiently large value of $n$. 

Our work has a potential application for generating certified device-independent randomness in the sequential scenario following the line of work developed in \cite{ran1}. Further studies along this line are thus called for. Finally, we note here that the preparation contextuality is a weaker correlation than the non-locality. It is worthwhile to explore the possibility of formulating suitable local realist inequality to investigate the sharing of nonlocality for more than one pair of sequential observers. It would also be interesting to formulate new preparation non-contextual inequality for multi-outcome and multi-party scenarios for demonstrating the sharing of preparation contextuality. Studies along this line could be an interesting avenue of research.

\section*{Acknowledgments}
 AKP acknowledges the support from the project DST/ICPS/QuST/Theme 1/Q42.


\begin{thebibliography}{99}
\bibitem{bell} J.S. Bell, On the Einstein Podolsky Rosen paradox, \href{https://doi.org/10.1103/PhysicsPhysiqueFizika.1.195}{Physics, {\bf 1}, 195 (1964)}.





\bibitem{bar05} J. Barrett, L. Hardy and A. Kent, No Signaling and Quantum
Key Distribution,\href{https://doi.org/10.1103/PhysRevLett.95.010503}{ Phys. Rev. Lett. 95, 010503(2005).}
\bibitem{acin06} A. Acin, N. Gisin and L. Masanes, From Bell’s Theorem to
Secure Quantum Key Distribution, \href{https://doi.org/10.1103/PhysRevLett.97.120405}{Phys. Rev. Lett. 97, 120405.}
(2006).
\bibitem{acin07} A. Acin, N. Brunner, N. Gisin, S. Massar, S. Pironio and and
V. Scarani, Device-Independent Security of Quantum Cryptography against Collective Attacks, \href{https://doi.org/10.1103/PhysRevLett.98.230501}{Phys. Rev. Lett. 98, 230501
(2007).}
\bibitem{pir09} S. Pironio, A. Acin, N. Brunner, N. Gisin, S. Massar and V.
Scarani, Device-independent quantum key distribution secure
against collective attacks, \href{https://doi.org/10.1088/1367-2630/11/4/045021}{New J. Phys. 11, 045021 (2009).}





\bibitem{col06} R. Colbeck, Quantum and relativistic protocols for secure
multi-party computation, Ph.D. thesis, University of Cambridge
(2006), \href{https://doi.org/10.48550/arXiv.0911.3814}{ arXiv:0911.3814v2.}


\bibitem{pir10} S. Pironio, et al., Random numbers certified by Bell's theorem,{Nature 464, 1021 (2010).}

\bibitem{nieto} O. Nieto-Silleras, S. Pironio and J. Silman, Using complete
measurement statistics for optimal device-independent randomness evaluation, {New J. Phys. 16, 013035 (2014).}

\bibitem{col12} R. Colbeck and R. Renner, Free randomness can be amplified,
\href{https://doi.org/10.1038/nphys2300}{Nature Phys. 8, 450 (2012).}

 


\bibitem{wehner} S. Wehner, M. Christandl and A.C. Doherty, Lower bound on
the dimension of a quantum system given measured data, \href{https://doi.org/10.1103/PhysRevA.78.062112}{Phys.
Rev. A 78,062112 (2008).}

\bibitem{gallego} R. Gallego, N. Brunner, C. Hadley and A. Acin, DeviceIndependent Tests of Classical and Quantum Dimensions, \href{https://doi.org/10.1103/PhysRevLett.105.230501}{Phys.
Rev. Lett. 105, 230501 (2010).}

\bibitem{ahrens} J. Ahrens, P. Badziag, A. Cabello and M. Bourennane, Experimental device-independent tests of classical and quantum dimensions, \href{https://doi.org/10.1038/nphys2333}{Nature Phys. 8, 592 (2012).}

\bibitem{brunnerprl13} N. Brunner, M. Navascues and T. Vertesi, Dimension Witnesses and Quantum State Discrimination, \href{https://doi.org/10.1103/PhysRevLett.110.150501}{Phys. Rev. Lett. 110,
150501 (2013).}
\bibitem{bowler} J. Bowles, M. Quintino and N. Brunner, Certifying the Dimension of Classical and Quantum Systems in a Prepare-and Measure Scenario with Independent Devices, \href{https://doi.org/10.1103/PhysRevLett.112.140407}{Phys. Rev. Lett.
112, 140407 (2014).}

\bibitem{sik16prl} J. Sikora, A. Varvitsiotis and Z. Wei, Minimum Dimension of
a Hilbert Space Needed to Generate a Quantum Correlation,
\href{https://doi.org/10.1103/PhysRevLett.117.060401}{Phys. Rev. Lett., 117, 060401 (2016).}
\bibitem{cong17} W. Cong, Y. Cai, J-D. Bancal and V. Scarani, Witnessing Irreducible Dimension, \href{https://doi.org/10.1103/PhysRevLett.119.080401}{Phys. Rev. Lett. 119, 080401 (2017).}
\bibitem{pan2020} A. K. Pan and S. S. Mahato, Device-independent certification of
the Hilbert-space dimension using a family of Bell expressions,
\href{https://doi.org/10.1103/PhysRevA.102.052221}{Phys. Rev. A 102, 052221 (2020).}
\bibitem{complx1} H. Buhrman, R. Cleve, S. Massar, and R. de Wolf, Nonlocality
and communication complexity, \href{https://doi.org/10.1103/RevModPhys.82.665}{Rev. Mod. Phys. 82, 665 (2010).}




\bibitem{kochen67} S. Kochen and E. Specker, The Problem of Hidden Variables in Quantum Mechanics, \href{https://www.jstor.org/stable/24902153}{ J. Math. Mech. 17, 59 (1967).}



\bibitem{peres90} A. Peres, Incompatible results of quantum measurements, \href{https://doi.org/10.1016/0375-9601(90)90172-K}{ Phys. Lett. A 151, 107 (1990).}

\bibitem{mermin93} N. D. Mermin, Hidden variables and the two theorems of John Bell, \href{https://doi.org/10.1103/RevModPhys.65.803}{ Rev. Mod. Phys. 65, 803 (1993).}
\bibitem{cab08} A. Cabello, Experimentally Testable State-Independent Quantum Contextuality, \href{https://doi.org/10.1103/PhysRevLett.101.210401}{Phys. Rev. Lett. 101, 210401 (2008).}
 \bibitem{pan10} A. K. Pan, A variant of Peres-Mermin proof for testing noncontextual realist models, \href{ https://doi.org/10.1209/0295-5075/90/40002}{EPL 90, 40002 (2010).}


	\bibitem{spek05} R. W. Spekkens, Contextuality for preparations, transformations, and unsharp measurements, \href{https://journals.aps.org/pra/abstract/10.1103/PhysRevA.71.052108}{Phys. Rev. A 71, 052108 (2005).}

 


	\bibitem{spek09} R. W. Spekkens,  D. H. Buzacott, A. J.  Keehn, B. Toner and G. J. Pryde,  Preparation contextuality powers parity-oblivious multiplexing, \href{https://journals.aps.org/prl/abstract/10.1103/PhysRevLett.102.010401}{Phys. Rev. Lett. 102, 010401 (2009)} .
    
    \bibitem{hameedi} A. Hameedi, A. Tavakoli, B. Marques, and M. Bourennane, Communication Games Reveal Preparation Contextuality, \href{https://journals.aps.org/prl/abstract/10.1103/PhysRevLett.119.220402}{Phys. Rev. Lett. 119, 220402 (2017)}.
	
	\bibitem{ghorai18} S. Ghorai and A. K. Pan, Optimal quantum preparation contextuality in an $n$-bit parity-oblivious multiplexing task, \href{https://journals.aps.org/pra/abstract/10.1103/PhysRevA.98.032110}{Phys. Rev. A 98, 032110 (2018)}.
	
	\bibitem{saha19b} D. Saha and A. Chaturvedi, Preparation contextuality as an essential feature underlying quantum communication advantage, \href{https://journals.aps.org/pra/abstract/10.1103/PhysRevA.100.022108}{Phys. Rev. A 100, 022108 (2019)}.
	
	\bibitem{pan19} A. K. Pan, Revealing universal quantum contextuality through communication games, \href{https://www.nature.com/articles/s41598-019-53701-5}{Sci Rep 9, 17631 (2019)}.

    
    
     \bibitem{schmid18}  D. Schmid and R. W. Spekkens, Contextual Advantage for State Discrimination, \href{https://journals.aps.org/prx/abstract/10.1103/PhysRevX.8.011015}{Phys. Rev. X 8, 011015 (2018)}.
     
      \bibitem{mukherjee22} S. Mukherjee, S. Naonit and A.K. Pan, Discriminating three mirror symmetric states with restricted contextual advantage, \href{https://journals.aps.org/pra/abstract/10.1103/PhysRevA.106.012216?ft=1}{Phys. Rev. A 106, 012216 (2022)}.
     
  

\bibitem{kumari2019}  A. Kumari and A.K. Pan, Sharing nonlocality and nontrivial preparation contextuality using the same family of Bell expressions, \href{https://journals.aps.org/pra/abstract/10.1103/PhysRevA.100.062130}{ Phys. Rev. A 100, 062130 (2019).}


			\bibitem{pan21} A. K. Pan, Oblivious communication game, self-testing of projective and nonprojective measurements, and certification of randomness, \href{https://journals.aps.org/pra/abstract/10.1103/PhysRevA.104.022212}{Phys. Rev. A, 104, 022212 (2021).}



	











\bibitem{clauser69}	J. F. Clauser, M. A. Horne, A. Shimony, and R. A. Holt,
Proposed Experiment to Test Local Hidden-Variable Theories, \href{https://doi.org/10.1103/PhysRevLett.23.880 }{
Phys. Rev. Lett. 23, 880 (1969)}.

\bibitem{silva2015} R. Silva, N. Gisin, Y. Guryanova and S. Popescu,
\href{https://journals.aps.org/prl/abstract/10.1103/PhysRevLett.114.250401}{Phys. Rev. Lett. 114, 250401 (2015).}

\bibitem{brown2020}  P. J. Brown and R. Colbeck, Arbitrarily Many Independent Observers Can Share the Nonlocality of a Single Maximally Entangled Qubit Pair, \href{https://journals.aps.org/prl/abstract/10.1103/PhysRevLett.125.090401}{Phys. Rev. Lett. 125, 090401  (2020)}.




\bibitem{Cheng2021} S. Cheng, L. Liu, T. J. Baker, and M. J. W. Hall, Limitations on sharing Bell nonlocality between sequential pairs of observers, 
\href{https://journals.aps.org/pra/abstract/10.1103/PhysRevA.104.L060201}{Phys. Rev. A. 104, L060201 (2021).}

\bibitem{Cheng2022} S. Cheng, L. Liu, T. J. Baker, and M. J. W. Hall, Recycling qubits for the generation of Bell nonlocality between independent sequential observers, 
\href{https://doi.org/10.1103/PhysRevA.105.022411}{Phys. Rev. A 105, 022411 (2022).}




\bibitem{bera2018}  A. Bera, S. Mal, A. Sen(De) and U. Sen, Witnessing bipartite entanglement sequentially by multiple observers, \href{https://journals.aps.org/pra/abstract/10.1103/PhysRevA.98.062304}{ Phys. Rev. A 98, 062304 (2018)}.


\bibitem{sasmal2018} S. Sasmal , D. Das , S. Mal and A. S. Majumdar, Steering a single system sequentially by multiple observers, \href{https://journals.aps.org/pra/abstract/10.1103/PhysRevA.98.012305}{Phys. Rev. A 98, 012305  (2018)}



\bibitem{Akshata} A. Shenoy, H. S. Designolle, F. Hirsch, R. Silva, N. Gisin, and N. Brunner, Unbounded sequence of observers exhibiting Einstein-Podolsky-Rosen steering,
\href{https://journals.aps.org/pra/abstract/10.1103/PhysRevA.99.022317}{Phys. Rev. A 99, 022317 (2019).}

\bibitem{Shashank} S. Gupta, A. G. Maity, D. Das, A. Roy, and A. S. Majumdar, Genuine Einstein-Podolsky-Rosen steering of three-qubit states by multiple sequential observers,
\href{https://journals.aps.org/pra/abstract/10.1103/PhysRevA.103.022421}{Phys. Rev. A 103, 022421 (2021).}


			
\bibitem{zh21} T. Zhang and S-M. Fei, Sharing quantum nonlocality and genuine nonlocality with independent observables, \href{https://journals.aps.org/pra/abstract/10.1103/PhysRevA.103.032216}{Phys. Rev. A 103, 032216  (2021)}


	
			

\bibitem{Karthik} K. Mohan, A. Tavakoli and N. Brunner, Sequential random access codes and self-testing of quantum measurement instruments,
\href{https://iopscience.iop.org/article/10.1088/1367-2630/ab3773}{New J. Phys. 21 083034 (2019).}
\bibitem{sumit} S. Mukherjee and A. K. Pan, Semi-device-independent certification of multiple unsharpness parameters through sequential measurements,
\href{https://journals.aps.org/pra/abstract/10.1103/PhysRevA.104.062214}{Phys. Rev. A 104, 062214(2021).}







\bibitem{roy2020} S. Roy, A. Kumari, S. Mal, A. Sen(De), Robustness of Higher Dimensional Non-locality against dual noise and sequential measurements,{	arXiv:2012.12200 (2020)}.

\bibitem{Mao2022} Y. L. Mao \emph{et al.}, Recycling non-locality in a quantum network,	
\href{https://arxiv.org/abs/2202.04840}{arXiv:2202.04840 (2022).}

\bibitem{anwer21} H. Anwer, N. Wilson, R. Silva, S. Muhammad, A. Tavakoli and M. Bourennane, Noise-robust preparation contextuality shared between any number of observers via unsharp measurements \href{	https://doi.org/10.22331/q-2021-09-28-551}{	Quantum 5, 551 (2021).}
			
			\bibitem{ambainis} A. Ambainis, D. Leung, L. Mancinska, M. Ozols, Quantum Random Access Codes with Shared Randomness, \href{https://arxiv.org/abs/0810.2937}{ arXiv:0810.2937v3 (2008)}. 
		
			
			
		
			
			\bibitem{gisin}	N. Gisin, Bell inequalities: many questions, a few answers,{ arXiv:quant-ph/0702021.}

			\bibitem{busch} P. Busch, Unsharp reality and joint measurements for spin observables, \href{https://journals.aps.org/prd/abstract/10.1103/PhysRevD.33.2253}{Phys. Rev. D. 33, 2253 (1986)}.


\bibitem{std1} J. Bergou, E. Feldman, and M. Hillery, Extracting Information from a Qubit by Multiple Observers: Toward a Theory of Sequential State Discrimination, \href{https://doi.org/10.1103/PhysRevLett.111.100501}{Phys. Rev. Lett. 111,
100501 (2013).}

\bibitem{std2} D. Fields, R. Han, M. Hillery and J. A. Bergou, Extracting unambiguous information from a single qubit by sequential observers, \href{https://doi.org/10.1103/PhysRevA.101.012118}{Phys. Rev. A 101, 012118 (2020).}

\bibitem{ran1} F. J. Curchod, M. Johansson, R. Augusiak, M. J. Hoban, P. Wittek, and A. Acın, Unbounded randomness certification using sequences of measurements, \href{https://doi.org/10.1103/PhysRevA.95.020102}{Phys. Rev. A 95, 020102(R) (2017).}











	
		
\end{thebibliography}
\end{document}